\documentclass[twocolumn, showpacs, preprintnumbers, amsmath, amssymb]{revtex4}
\usepackage{graphicx}% Include figure files
\usepackage{dcolumn}% Align table columns on decimal point
\usepackage{bm}% bold math
\usepackage{color}
\begin{document}
%--------------------------------------------------------------------
\title{Self-Affinity in the Gradient Percolation Problem}
%--------------------------------------------------------------------

\author{Alex Hansen}
\email[Alex.Hansen@ntnu.no]{}
\affiliation{Department of Physics, Norwegian University of Science and
Technology, N--7491 Trondheim, Norway}

\author{G.\ George Batrouni} \email[George.Batrouni@inln.cnrs.fr]{}
\altaffiliation{INLN, UMR CNRS 6618, Universit{\'e} de Nice-Sophia
Antipolis, 1361 route des Lucioles, F--06560 Valbonne, France}
\affiliation{Department of Physics, Norwegian University of Science
and Technology, N--7491 Trondheim, Norway}

\author{Thomas Ramstad}
\email[Thomas.Ramstad@phys.ntnu.no]{}
\affiliation{Department of Physics, Norwegian University of Science and
Technology, N--7491 Trondheim, Norway}

\author{Jean Schmittbuhl}
\email[Jean.Schmittbuhl@eost.u-strasbg.fr]{} \affiliation{Institut de
Physique du Globe de Strasbourg, UMR CNRS 7516, 5, rue Ren{\'e}
Descartes, F--67084 Strasbourg, France}

%--------------------------------------------------------------------
\date{\today}
%--------------------------------------------------------------------
\begin{abstract}
We study the scaling properties of the solid-on-solid front of the
infinite cluster in two-dimensional gradient percolation. We show that
such an object is self affine with a Hurst exponent equal to 2/3 up to
a cutoff-length $\sim g^{-4/7}$, where $g$ is the gradient. Beyond
this length scale, the front position has the character of
uncorrelated noise. Importantly, the self-affine behavior is robust
even after removing local jumps of the front. The previously observed
multi affinity, is due to the dominance of overhangs at small
distances in the structure function.  This is a crossover effect.
\end{abstract}
%--------------------------------------------------------------------
%--------------------------------------------------------------------
\maketitle
%--------------------------------------------------------------------
Rough surfaces showing non-trivial scaling properties have been
extensively studied theoretically, numerically and experimentally over
the last couple of decades.  Examples of such surfaces are those
appearing during brittle fracture \cite{b97} which were first
characterized as being fractal \cite{mpp84-bs85} but it was then
realized that the concept of self affinity was more appropriate
\cite{blp90-mhhr92}.  The question of self affinity versus fractality
has also been the focus of intense research on invasion fronts in
porous media and on the dynamics of magnetic domain walls
\cite{cr88-cr90-mcr91-jr92}.  It was recently reported that the
displacement fronts in self-affine fractures are self affine
\cite{dakh04}.  More recently, a possible explanation for the observed
self affinity of fracture surfaces has been proposed and hinges on a
clear understanding of the distinction between fractality and self
affinity \cite{hs03a-hs03b,az04}. It has also been suggested that
brittle fracture surfaces are multi affine rather than simply self
affine \cite{sss95-bpsv06}.  Whether this is so remains an open
question \cite{smmhsvdbr06}.

%--------------------------------------------------------------------
\begin{figure}[b]
\includegraphics[width=7cm,clip]{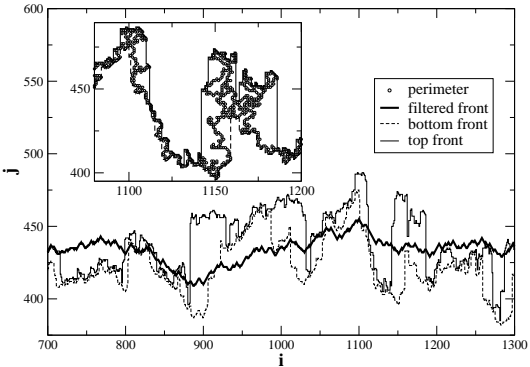}
\caption{Top-side and bottom-side SOS fronts based on
the perimeter of the cluster connected to the $p=1$ edge (shown in
the insert as dots).  We also show the filtered $j_0(i)$ front 
(see Eq.\ (\ref{transform})).}
\label{fig1}
\end{figure}
%--------------------------------------------------------------------

It is the aim of this Letter to study the question of fractality, self
affinity and multi affinity of a front in a system which is simple
enough to be tractable, namely that of the gradient percolation
\cite{srg85}.  There are already in the literature studies of this
system in the present context. Furuberg {\it et al.\/} \cite{fhhfj91} study
the jumps in the position of the solid-on-solid (SOS) front of the
infinite cluster, whereas Asikainen {\it et al.\/} \cite{amda02} conclude that
this front is multi affine.  We will in this Letter show that up to a
given scale, the SOS front is {\it self affine\/} with a well-defined
Hurst exponent, whereas on larger scales its position becomes
uncorrelated. The self affinity is {\it not\/} caused by the jumps in
the position of the front due to overhangs, but related to its fractal
structure.  The multi affinity seen by Asikainen {\it et al.\/} has its
origin in the overhangs resulting from the definition of the SOS
fronts and shows up in the structure function on small scales.

In {\it gradient\/} percolation, a spatial gradient in the occupation
probability $p$ is introduced.  A cartesian coordinate system $(i,j)$
is oriented with respect to the finite lattice of size $L_i\times L_j$
(assuming for the rest of this paper that the lattice is two
dimensional), so that the $i$ axis runs perpendicular to the gradient
(i.e.\ along the lower edge) and the $j$ axis along the gradient
(i.e.\ the left edge). The gradient is introduced in the $j$ direction
so that $p(j)=gj$, where the gradient $g=1/L_j$.  However, the cluster
connected to the lower edge will reach some average value, $j=j_g$,
with an associated occupation probability $p_g=gj_g$.  The region
around $j_g$ is critical and has a width $\xi$, spanning between
$j_{\pm}=j_g\pm\xi/2$, where $\xi$ is the correlation length
associated with the critical region in the direction of the gradient.
Defining $p_{\pm}=gj_{\pm}$ and setting
$\xi=|p_{\pm}-p_g|^{-\nu}=|g(j_{\pm}-j_g)|^{-\nu}$ where $\nu$ is the
correlation length exponent, Sapoval {\it et al.\/} \cite{srg85} found that
$\xi\sim g^{-\nu/(1+\nu)}=g^{-4/7}$.

%--------------------------------------------------------------------
\begin{figure}[t]
\includegraphics[width=7cm,clip]{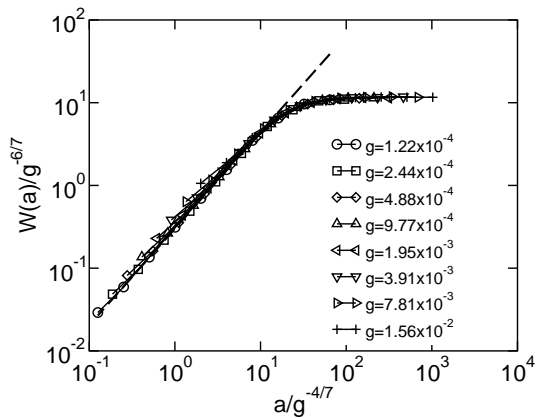}
\caption{Data collapse of the averaged wavelet coefficients for the
bottom side front based on lattice size $L_j=64$ to 8192, while
$L_i=2048$.  We have that $g=1/L_j$. The straight line has a slope of
$\zeta+1/2$, see Eq.\ (\ref{awc}).}
\label{fig2}
\end{figure}
%--------------------------------------------------------------------

The infinite cluster has a fractal structure with an upper cutoff in
length scale set by the width of the critical region, $\xi$.  We now
focus on the front of this infinite cluster and define precisely what
we mean by this front $j(i)$ in the gradient percolation problem.  Our
starting point is the perimeter of the cluster of occupied sites that
is attached to the $p=1$ edge of the lattice.  Since this perimeter
contains overhangs and therefore is multivalued when interpreted as a
function $j(i)$, we use the SOS method to extract a single-valued
function for its position, see Fig.\ \ref{fig1}.  For each $i$, we use
either the $j$ value that is closest to the $p=0$ edge (top side) or
the $j$ value which is closest to the $p=1$ side (bottom side) or the
average over all the $j$ values attached to a given $i$ value (average
front).

A trace $j(i)$ is statistically self affine if the probability
density, $\pi(i,j)$, for it to have a value $j$ at $i$, given that
$j=0$ at $i=0$, has the invariance
\begin{equation}
\label{selfaffine}
\lambda^\zeta \pi (\lambda i,\lambda^\zeta j)=\pi(i,j)\;,
\end{equation}    
where $\zeta$ is the Hurst exponent.  This invariance must be caused
by {\it spatial correlations} in $j$ along the $i$-axis.  We note that
a L{\'e}vy flight, which is an uncorrelated random walk whose step
size $h$ is drawn from a power law distribution $N(h)\sim
h^{-\beta-1}$, will satisfy Eq.\ (\ref{selfaffine}) with an apparent
Hurst exponent $\zeta=1/\beta$.  However, in this example, satisfying
Eq.\ (\ref{selfaffine}) is due to the step size distribution and not
to spatial correlations \cite{hm06}.

We have used the Average Wavelet Coefficient (AWC) method
\cite{mrs97,shn98} to analyse the structure of the SOS fronts.  The
AWC method consists of wavelet transforming $j(i)$, and averaging
the wavelet coefficients $w(b,a)$ at each length scale $a$ over
position $b$, $W(a)=\langle w(b,a)\rangle_b$.  If $j(i)$ is self
affine, the averaged wavelet coefficients will scale as
\begin{equation}
\label{awc}
W(a) \sim a^{\zeta+1/2}\;.
\end{equation} 
We show in Fig.\ \ref{fig2}, the averaged wavelet coefficients based
on the Daubechies-4 wavelets for the bottom-side fronts.  The plots
for the top and average fronts are comparable.
The data are based on averages over
2201 samples for $L_j$ in the range 64 to 2048 and 200 samples for
$L_j=4096$ and 8192. $L_i$ was set to 2048 for all the different
$L_j$.  The gradient $g$ was set to $1/L_j$. There is a clear
crossover between two regimes in these plots.  At smaller length
scales, one does indeed find the behavior of Eq.\ (\ref{awc})
indicating self affinity.  On larger scales, the slope of the log-log
plots are zero indicating $\zeta=-1/2$, which corresponds to
uncorrelated or white noise \cite{hsb01}.  Furthermore, we observe
excellent data collapse when $W$ is scaled by $g^{-\beta}$ and the length
scale, $a$, is scaled by $g^{-\alpha}$.  We will show below that
\begin{equation}
\label{zeta}
\zeta=2-D_e =\frac{2}{3}\;,
\end{equation} 
\begin{equation}
\label{alpha}
\alpha=\frac{\nu}{1+\nu}=\frac{4}{7}\;,
\end{equation}
and
\begin{equation}
\label{beta}
\beta=\frac{3}{2}\ \alpha=\frac{6}{7}\;.
\end{equation}
where $D_e=4/3$ is the fractal dimension of the external perimeter of the 
front \cite{ga86-ga87}.

%--------------------------------------------------------------------
\begin{figure}[t]
\includegraphics[width=7cm,clip]{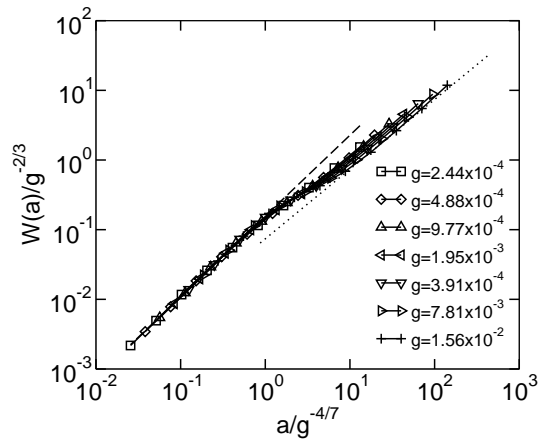}
\caption{Data collapse of the averaged wavelet coefficients for the
smoothed $j_0(i)$ based on the bottom side front.  The lattice sizes
and gradients are as in Fig.\ \ref{fig2}.  The long-dashed line has a
slope of 7/6, see Eq.\ (\ref{awc}), whereas the dotted line has a
slope of 1=1/2+1/2, consistent with uncorrelated random walks.}
%Note the change of scaling along the $j$ axis.}
\label{fig9}
\end{figure}
%--------------------------------------------------------------------

The main goal of this Letter is to derive Eq.\ (\ref{zeta}) and thus
demonstrate that $\zeta$ is a proper Hurst exponent and $j(i)$ a
self-affine function.  To this end, we need to demonstrate two things:
First, $j(i)$ satisfies the scaling relation (\ref{selfaffine}) and,
second,that this is not due to a power law tail in the step size
distribution.  We note that since the average wavelet coefficients
obey Eq.\ (\ref{awc}), $j(i)$ automatically satisfies Eq.\
(\ref{selfaffine}).  Therefore, we now need only to identify the
mechanism behind this scaling.

In order to derive Eq.\ (\ref{zeta}), we start by noting that the
distribution of distances $m$ between crossing points between a planar
fractal curve with dimension $D_e$ --- e.g., the percolation perimeter
--- and a straight line follows the power law $\pi(m)\sim m^{-D_e}$
\cite{hme94}.  Introducing a gradient in the $j$ direction and placing
the straight line in the critical region interval, $[j_-,j_+]$, and
parallel to the $i$ axis, the distribution of crossing point distances
$m$ remains the same.  A self-affine curve characterized by a Hurst 
exponent $\zeta$, leads to a distribution of crossing point
distances given by $\pi(m)\sim m^{-(2-\zeta)}$ \cite{hme94}.  By comparing
this expression to $\pi(m)\sim m^{-D_e}$,  
Eq.\ (\ref{zeta}) immediately follows.  However, we still need to show
that $j(i)$ is indeed self affine, in other words the scaling relation
Eq.\ (\ref{selfaffine}) is not caused by jumps.

First, we turn to deriving Eqs.\ (\ref{alpha}) and (\ref{beta}).  The
correlation length in the direction of the gradient, the $j$
direction, is $\xi\sim g^{-\nu/(1+\nu)}$.  Since the perimeter is
locally isotropic, this is also the correlation length in the $i$
direction.  The crossover length scale from self affinity to
uncorrelated noise is the correlation length $\xi$.  Hence, rescaling
$a \to a/\xi\sim a/g^{-\nu/(1+\nu)}$ gives data collapse along this
axis which demonstrates Eq.\ (\ref{alpha}).  Likewise, the crossover
length scale in the $j$ direction is $\xi$.  This implies that the
normalized wavelet coefficient at this scale, $W(\xi)/\xi^{1/2}$ is
equal to $\xi$.  Hence, $W(\xi)\sim \xi^{3/2} \sim g^{-(3/2)\alpha}
\sim g^{-\beta}$, and $\beta=(3/2)\alpha= 6/7$, as stated in Eq.\
(\ref{beta}).

In order to show that $j(i)$ is a self-affine function, we need to
demonstrate that $\zeta$ is not caused by the step size distribution.
To this end, we define the following transformation of the function
$j(i)\to j_k(i)$ where we factorize the function in such away that we
can distinguish the respective roles of persistency and step sizes,
\begin{equation}
\label{transform}
j_k(i)=\sum_{m=0}^i {\rm sign}[j(m+1)-j(m)]\ |j(m+1)-j(m)|^k\;,
\end{equation}
where $|j(m+1)-j(m)|=h(m)$ is the step size at position $m$.
We have in particular that $j_1(i)=j(i)$.  
It was shown in \cite{fhhfj91}, that $h$ is distributed according to 
\begin{equation}
\label{overhang}
N(h,g)=h^{-D_e-1}f(hg^\alpha)\;,
\end{equation}
where $D_e=4/3$ and $f(z)$ approaches a constant as $z\ll 1$ and falls
off faster than any power law as $z\to\infty$.  The step size
distribution comes from the appearance of overhangs in the perimeter.
An overhang is defined as the jump made by the front from one position
along the $i$ axis to the next due to a backwards turn
\cite{hahw90,fhhfj91,bh92}.  In order to confirm that the overhangs do
not generate the Hurst exponent $\zeta=2/3$, we analyse the filtered
front $j_0(i)$, defined in Eq.\ (\ref{transform}). With $k=0$, we
eliminate the overhangs all together \cite{bmr05}.  Fig.\
\ref{fig9} shows the data collapse based on $j_0(i)$ corresponding to 
the bottom side $j(i)$ shown in Fig.\ \ref{fig2}.  The scaling 
along the $i$ axis
is unchanged as no change in the system has been made in that
direction.  However, since all step sizes have been reset to unity in
the transformation $j(i)\to j_0(i)$, the rescaling in the $j$
direction is no longer controlled by $j_c$.  In order to regain data
collapse for different $g=1/L_j$, we need to rescale the lattice units
in this direction by the Hurst exponent, $\zeta=2/3$.  The straight
line matching the small-$a$ region of the figure has a slope
$2/3+1/2$, while the straight line matching the large-$a$ portion has
a slope of $1/2+1/2$ corresponding to an uncorrelated random walk.
This shows that, for small scales, the $\zeta=2/3$ is indeed a Hurst
exponent. On the other hand, for longer length scales, we expect
random walk behavior since white noise gives precisely the exponent
$1/2$ in the transformation $j(i)\to j_0(i)$.

\begin{figure}[t]
\includegraphics[width=7cm,clip]{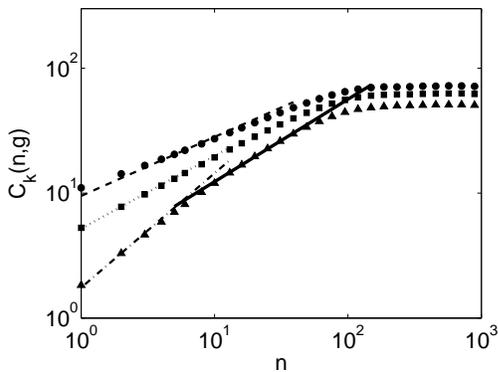}
\caption{$C_k(n,g)$ as a function of $n$ for $k=1$, 2 and 3.  The
three leftmost straight lines have slopes according to Eq.\ (\ref{affen}),
while the bold middle line has a slope equal to 2/3 in accordance with 
Eq.\ (\ref{chimps}). For large $n$ the structure functions become flat
indicating that one has reached the decorrelated regime.}
\label{fig10}
\end{figure}

In order to analyse the multi affinity that has been reported in this
problem \cite{amda02}, we construct the structure function
$C_k(n,g)=\langle |j(m+n)-j(m)|^k\rangle$.  Multi affinity occurs when
$C_k(n,g)^{1/k}$ does {\it not\/} scale with a single $k$-independent
exponent with respect to $n$.  Using the overhang distribution
(\ref{overhang}), we find $C_k(1,g)\sim g^{s(k)}$, where $s(k)
=\min[0,\alpha(D_e-k)]=\min[0,(16/21-4 k/7)]$.  The self-affine
character of $j(i)$ cannot be visible in the structure function for
$n=1$ but will appear only gradually as $n$ is increased.  We may
therefore analyse the structure function based solely on the L{\'e}vy
character induced by the overhangs in the small-$n$ limit.  We will
call this the L{\'e}vy regime, whereas for larger $n$ where the self
affinity dominates, we will refer to as the self-affine regime.  The
scaling with respect to $g$ for $C_k(1,g)$ persists for $n>1$ in the
L{\'e}vy regime since $j(i+n)-j(i)$ follows a L{\'e}vy distribution
whose power law tail does not change with increasing $n$.  Hence, we
expect $C_k(n,g)\sim g^{s(k)}$ in this regime.  In order to derive its
dependence on $n$ in the L{\'e}vy regime, we note that the
distribution of distances $l$ between overhangs follows the
same power law as the overhangs themselves. This can be seen as
follows.  When there is a gradient present in the $j$ direction, the
length of the perimeter scales as $L_i^{D_e}$, when the gradient is
kept fixed.  Making a cut through the perimeter with a straight line
parallel to the $i$ axis, the crossing points of the perimeter with
the line form a fractal set with dimension $D_e-1$. Hence, there are,
in a given interval $l$, $N_l \sim l^{D_e-1}$ overhangs \cite{hme94}.
These overhangs give rise to an effective Hurst exponent $1/D_e=3/4$
on the fractal set, seen e.g.\ in the width of the trace, $\Delta j
\sim N_l^{1/D_e}\sim l^{(D_e-1)/D_e}$.  Since the overhangs form a
fractal set, we will need $N_b \sim l^{-(D_e-1)}$ boxes of size $l$ to
cover it.  Due to the averaging over position $i$, there will be yet
another factor $l$, see \cite{m05}.  We may now assemble these pieces
to form the scaling of the structure function in the L{\'e}vy regime,
$C_k(l,g)\sim N_l^k N_b l \sim l^{k\zeta_k^L}$ where
\begin{equation}
\label{affen}
\zeta_k^L= \left[1-\frac{1}{D_e}\right]+\frac{2-D_e}{k} = \frac{1}{4}
+\frac{2}{3k}\;.
\end{equation} 
Therefore, in the L{\'e}vy regime, i.e.\ for small $n$, there is multi
affinity. A similar analysis in the self-affine regime, i.e.\ at
larger $n$, yields  
\begin{equation}
\label{chimps}
\zeta_k^{SA} = \zeta = \frac{2}{3}\;.
\end{equation}
Therefore there is no multi affinity in this regime.
The $n$ for which there is the 
crossover between the L{\'e}vy and the self-affine regime will depend on
$k$ and is governed by prefactors that the scaling analysis presented here
cannot access.  For $n$ beyond $\xi$,  the front decorrelates
and the structure function becomes independent of $n$.
We show in Fig.\ \ref{fig10}, the $k=1$, 2 and 3 structure functions.  Their
behavior is in accordance with our predictions. However, note that for $k=2$,
$\zeta_2^L=7/12=0.58$ which is close to $\zeta=2/3$. Furthermore, the 
self-affine regime is close to the decorrelated flat regime.  Hence, it is 
hard to distinguish between the L{\'e}vy and the self-affine regime for this
value of $k$.  As $k$ increases, the L{\'e}vy regime grows, as the overhangs
are emphasized for larger $k$. 

%--------------------------------------------------------------------
%--------------------------------------------------------------------

To conclude, we have shown that the structure of the interface in a
gradient percolation problem combines fractal and self-affine
properties. The perimeter that includes numerous overhangs has the
classical fractal structure \cite{srg85}. However, Solid-on-Solid
fronts that are extracted from the perimeter, have a clear self-affine
property up to a crossover length scale $\xi$ even if local jumps,
inherited from overhangs, are removed. On larger scales it shows an
uncorrelated noise behavior. The structure function is, however,
sensitive to the overhangs on smaller scales and this implies a
multi-affine scaling behavior in this regime.  Implications of our
results for physical interpretations of analogical and numerical
experiments are important.

We thank M.\ K.\ Alava and S.\ Zapperi for stimulating comments leading 
to this work. 
G.\ G.\ B.\ thanks NTNU and Norsk Hydro for his appointment as the Lars 
Onsager Professor for 2004.
%---------------------------------------------------------------------

%---------------------------------------------------------------------

\begin{thebibliography}{99}

\bibitem{b97} E.\ Bouchaud, J.\ Phys.\ Condens.\ Matt.\ {\bf 9}, 
4319 (1997).
{\bf 55}, 349 (2006). 

\bibitem{mpp84-bs85} B.\ B.\ Mandelbrot, D.\ E.\ Passoja, and A.\ J.\
Paullay, Nature, {\bf 308}, 721 (1984); S.\ R.\ Brown and C.\ H.\
Scholz, J.\ Geophys.\ Res.\ {\bf 90}, 12575 (1985).

\bibitem{blp90-mhhr92} E.\ Bouchaud, G.\ Lapasset,
and J.\ Plan{\'e}s, Europhys.\ Lett.\ {\bf 13}, 73 (1990); K.\ J.\
M{\aa}l{\o}y, A.\ Hansen, E.\ L.\ Hinrichsen, and S.\ Roux, Phys.\
Rev.\ Lett.\ {\bf 68}, 213 (1992). 
(1995).

\bibitem{cr88-cr90-mcr91-jr92} M.\ Cieplak and M.\ O.\ Robbins, Phys.\ Rev.\
Lett.\ {\bf 60}, 2042 (1988); M.\ Cieplak and M.\ O.\ Robbins, Phys.\ Rev.\ B,
{\bf 41}, 11508 (1990); N.\ Martys, M.\ O.\ Robbins and M.\ Cieplak,
Phys.\ Rev.\ B {\bf 44}, 12294 (1991); H.\ Ji and M.\ O.\ Robbins, Phys.\ 
Rev.\ B,{\bf 46}, 14519 (1992).

\bibitem{dakh04} G.\ Drazer, H.\ Auradou, J.\ Koplik and J.\ P.\ Hulin,
Phys.\ Rev.\ Lett.\ {\bf 92}, 014501 (2004). 

\bibitem{hs03a-hs03b} 
A.\ Hansen and J.\ Schmittbuhl, Phys.\ Rev.\ Lett.\ {\bf 90},
045504 (2003); J.\ Schmittbuhl, A.\ Hansen and G.\ G.\ Batrouni, Phys.\
Rev.\ Lett.\ {\bf 90}, 045505 (2003).

\bibitem{az04} M.\ J.\ Alava and S.\ Zapperi, Phys.\ Rev.\ Lett.\ {\bf 92},
049601 (2004); J.\ Schmittbuhl, A.\ Hansen and G.\ G.\ Batrouni, Phys.\ Rev.\
Lett.\ {\bf 92}, 049602 (2004).

\bibitem{sss95-bpsv06} J.\ Schmittbuhl, F.\ Schmitt and C.\ Scholz, J.\ 
Geophys.\ Res.\ {\bf 100}, 5953 (1995); E.\ Bouchbinder, I.\ Procaccia, 
S.\ Santucci and L.\ Vanel, Phys.\ Rev.\ Lett.\ {\bf 96}, 055509 (2006). 

\bibitem{smmhsvdbr06} S.\ Santucci, J. Mathiesen, K.\ J.\ M{\aa}l{\o}y,
A.\ Hansen, J.\ Schmittbuhl, L.\ Vanel, A.\ Delaplace, J.\ {\O}.\ H.\ Bakke
and P.\ Ray, Cond-mat/0607385.

\bibitem{srg85} B.\ Sapoval, M.\ Rosso and J.\ F.\ Gouyet, J.\ Phys.\ Lett.\
(France) {\bf 46}, L149 (1985).

\bibitem{fhhfj91} L.\ Furuberg, A.\ Hansen, E.\ L.\ Hinrichsen, J.\ Feder and 
T.\ J{\o}ssang, Phys.\ Script.\ T {\bf 38}, 91 (1991).

\bibitem{amda02} J.\ Asikainen, S.\ Majaniemi, M.\ Dub{\'e} and T.\ 
Ala-Nissil{\"a}, Phys.\ Rev.\ E {\bf 65}, 052104 (2002).

\bibitem{hahw90} A.\ Hansen, T.\ Aukrust, J.\ M.\ Houlrik and I.\ Webman,
 J.\ Phys.\ A {\bf 23}, L145 (1990).

\bibitem{bh92} G.\ G.\ Batrouni and A.\ Hansen, J.\ Phys.\ A {\bf 25},
L1059 (1992).

\bibitem{hm06} A.\ Hansen and J.\ Mathiesen in {\it Modelling critical and
catas\-trophic phenomena in geoscience: A statistical phy\-sics approach,\/}
edited by P.\ Bhattacharyya and B.\ K.\ Chakrabarti (Springer Verlag,
Berlin, 2006). 

\bibitem{shb03} J.\ Schmittbuhl, A.\ Hansen and G.\ G.\ Batrouni, Phys.\
Rev.\ Lett.\ {\bf 90}, 045505 (2003).

\bibitem{mrs97} A.\ R.\ Mehrabi, H.\ Rassamdana and M.\ Sahimi,
Phys.\ Rev.\ E, {\bf 56}, 712 (1997). 

\bibitem{shn98} I.\ Simonsen, A.\ Hansen and O.\ M.\ Nes, Phys.\ Rev.\ E
{\bf 58}, 2779 (1998).

\bibitem{hsb01} A.\ Hansen, J.\ Schmittbuhl and G.\ G.\ Batrouni, Phys.\
Rev.\ E {\bf 63}, 062102 (2001).

\bibitem{hme94} A.\ Hansen, K.\ J.\ M{\aa}l{\o}y and T.\ Eng{\o}y,
Fractals, {\bf 2}, 527 (1994).

\bibitem{ga86-ga87} T.\ Grossmann and A.\ Aharony, J.\ Phys.\ A {\bf 19}, L745
(1986); T.\ Grossmann and A.\ Aharony, J.\ Phys.\ A {\bf 20}, L1193
(1987).

\bibitem{bmr05} G.\ M.\ Buend{\'\i}a, S.\ J.\ Mitchell and P.\ A.\ Rikvold,
Microelectronics J.\ {\bf 36}, 913 (2005).

\bibitem{m05} S.\ J.\ Mitchell, Phys.\ Rev.\ E, {\bf 72}, 065103(R) (2005).

\end{thebibliography}
\end{document}